\documentclass{PoS}
\newcommand{\figref}{Fig.~\ref}
\newcommand{\f}{\ensuremath{f_{\text{BBH}}^{\nu}}}

\newcommand{\egw}{\text{3~M}\ensuremath{_{\odot}\ }}
\newcommand{\egwerr}{\ensuremath{3^{+0.5}_{-0.5}}\text{~M}\ensuremath{_{\odot}\ }}
\newcommand*\diff{\mathop{}\!\mathrm{d}}
\newcommand{\figwidth}{.7}

\usepackage{siunitx}
\usepackage[utf8]{inputenc}

\graphicspath{{../figures/}}

\title{Constraints and prospects on gravitational wave and neutrino emission using GW150914}

\ShortTitle{Constraints and prospects on gravitational wave and neutrino emission using GW150914}

\author{Krijn D. de Vries\\
        IIHE/ELEM, Vrije Universiteit Brussel, Pleinlaan 2, 1050 Brussels, Belgium}

\author{\speaker{Gwenha\"el de Wasseige}\\
        IIHE/ELEM, Vrije Universiteit Brussel, Pleinlaan 2, 1050 Brussels, Belgium\\
         E-mail: \email{gdewasse@vub.ac.be}}

\author{Jean-Marie Fr\`ere\\
        Universit\'e Libre Bruxelles, Boulevard de la Plaine 2, 1050 Brussels, Belgium}

\author{Matthias Vereecken\\
        IIHE/ELEM, Vrije Universiteit Brussel, Pleinlaan 2, 1050 Brussels, Belgium\\
		 TENA, Vrije Universiteit Brussel, Intern. Solvay institutes, Pleinlaan 2, 1050 Brussels, Belgium}
\abstract{The recent LIGO observation of gravitational waves from a binary black hole merger triggered several follow-up searches from both electromagnetic wave as well as neutrino observatories.
Since in general, it is expected that all matter has been removed from the binary black hole environment long before the merger, no neutrino emission is expected from such mergers. Still, it remains interesting to test this hypothesis.
The ratio of the energy emitted in neutrinos with respect to gravitational waves represents a useful parameter to constrain the environment of such astrophysical events. In addition to putting constraints by use of the non-detection of counterpart neutrinos, it is also possible to consider the diffuse neutrino flux measured by the IceCube observatory as the maximum contribution from an extrapolated full class of BBHs.  Both methods currently lead to similar bounds on the fraction of energy that can be emitted in neutrinos. Nevertheless, combining both methods should allow to strongly constrain the source population in case of a future neutrino counterpart detection. The proposed approach can and will be applied to potential upcoming LIGO events, including binary neutron stars and black hole-neutron star mergers, for which a neutrino counterpart is expected.}

\FullConference{35th International Cosmic Ray Conference\\
		 10-20 July, 2017\\
		 Bexco, Busan, Korea}

\begin{document}



\section{Introduction}

With the first direct detection of gravitational waves, a new observational window on the universe has opened. Following the multimessenger paradigm, gravitational waves can provide additional information about known astrophysical objects or reveal those that are otherwise unobservable.

The first detected event, GW150914~\cite{Abbott:2016blz}, has a signal compatible with the merger of a black hole binary. In this event, two black holes, with masses equal to 36$^{+5}_{-4}$ M$_ {\odot}$ and 29$^{+4} _{-4}$ M$_ {\odot}$, merged to form a single black hole of mass 62$^{+4}_{-4}$ M$_\odot$, releasing \egwerr of energy into gravitational waves. A black hole binary system is expected to be free of all matter by the time of the merger~\cite{Perna:2016jqh}. Therefore, no emission is expected from such a system except for gravitational waves. Now that a binary black hole (BBH) merger has been detected, this expectation can be tested.

The detection of GW150914 triggered multiple follow-up searches for electromagnetic~\cite{Abbott:2016gcq} and neutrino~\cite{Adrian-Martinez:2016xgn, Abe:2016jwn, Aab:2016ras} signals. While none of these searches detected a significant signal, 
Fermi GBM did see a sub-threshold event resembling a short gamma-ray burst (GRB)~\cite{Connaughton:2016umz}. However, INTEGRAL imaged the same area and found no such signal~\cite{Savchenko:2016kiv}, casting doubt on the interpretation of this observation. Still, 
several models were constructed that could explain a possible short GRB from BBH mergers (e.g.~\cite{Perna:2016jqh,Bartos:2016dgn,Kotera:2016dmp}).

Given that Megaton-scale neutrino detectors such as IceCube~\cite{Aartsen:2016nxy}
are available, it is appropriate to test whether BBH mergers emit neutrinos. In the following, we will use the non-detection of signal neutrinos associated with GW150914~\cite{Adrian-Martinez:2016xgn} to put a generic bound on the high energy neutrino emission from BBH mergers, relative to the energy emitted in gravitational waves~\cite{deVries:2016ljw}. This is compared to the current bound coming from the diffuse astrophysical neutrino flux detected by IceCube~\cite{Aartsen:2013jdh}. Afterwards, we show how to interpret this bound in terms of specific models. Using this interpretation, we estimate the amount of neutrino emission that could be expected in realistic models for BBH mergers. Conversely, from the current limit on the neutrino emission, we obtain a bound on the amount of matter present in the environment of BBH mergers.

\section{Neutrino emission}
\label{section:theory}

The neutrino emission coming from (a class of) BBH mergers will be parametrised by the neutrino emission fraction
\begin{equation}
\f = \frac{E_\nu}{E_\mathrm{GW}},
\label{eq:fractiondef}
\end{equation}
where $E_\nu$ is the total amount of energy emitted in neutrinos, while $E_\mathrm{GW}$ is the total energy released in gravitational waves, given by LIGO. This \f\ can be probed directly by experiments searching for neutrino emission coincident with the merger event.

In addition, \f\ is constrained by the diffuse high energy astrophysical neutrino flux detected by IceCube~\cite{Aartsen:2013jdh}, since BBH mergers have been happening throughout the history of the universe. We calculate this bound following the approach in ~\cite{Kowalski:2014zda,Ahlers:2014ioa}, assuming that BBH mergers emit neutrinos following an $E^{-2}$-spectrum. The diffuse neutrino flux due to (a class of) BBH mergers occurring with a rate $R$~(in Gpc$^{-3}$~yr$^{-1}$) in the local universe is given by~\cite{Waxman:1998yy},
\begin{equation}
E^2 \left.\frac{\diff N_\nu}{\diff E_\nu}\right|_{\mathrm{obs}} = \left( \f t_H \frac{c}{4\pi} \xi_z\right) E^2 \left.\frac{\diff \dot{N}_\nu}{\diff E_\nu}\right|_{\mathrm{inj}, \f=1},
\label{eq:injtoflux}
\end{equation}
with the total injected flux
\begin{equation}
E^2 \left.\frac{\diff \dot{N}_\nu}{\diff E_\nu}\right|_{\mathrm{inj}, \f=1} = R\ E^2 \phi(E_\nu).
\label{eq:injflux}
\end{equation}
Here, $\phi(E_\nu)$ is the neutrino flux from a single BBH merger in case that $\f=1$,where the integrated neutrino energy is equal to $E_\mathrm{GW}$. The cosmic evolution of the sources is contained in $\xi_z$. For a merger rate following the star formation rate (SFR)~\cite{Hopkins:2006bw,Yuksel:2008cu}, one has $\xi_z \approx 2.4$.

The rate of BBH mergers in the local universe is determined by LIGO. When accounting for different possible mass distributions, the 90\% credible interval for this rate is given by~\cite{TheLIGOScientific:2016pea}
\begin{equation}
R = \SIrange{9}{240}{Gpc^{-3} yr^{-1}}.
\label{eq:LIGOrate}
\end{equation}
For BBH mergers with black holes similar to GW150914, the corresponding rate is given by~\cite{TheLIGOScientific:2016pea},
\begin{equation}
R_{\mathrm{GW150914}} = 3.4^{+8.8}_{-2.8} \mathrm{Gpc}^{-3} \mathrm{yr}^{-1}.
\label{eq:rate3m}
\end{equation}

The diffuse neutrino emission from BBH mergers (Eq.~\ref{eq:injtoflux}) is constrained to be below the diffuse astrophysical neutrino flux measured by IceCube~\cite{Aartsen:2015zva}
\begin{equation}
E^2 \Phi(E) = 0.84 \pm 0.3\times\SI{.e-8}{GeV\;cm^{-2} s^{-1} sr^{-1}}.
\label{eq:icdiff}
\end{equation}
Therefore, by combining information from neutrino and gravitational wave observatories, the astrophysical bound constrains the source evolution (through $\xi_z$) and environment (through \f).

\section{Limits from GW150914}
\label{section:direct}

Using the null detection of counterpart neutrinos to GW150914 by IceCube~\cite{Adrian-Martinez:2016xgn}, we calculate the bound on the neutrino emission fraction \f. Afterwards, we compare this with the constraint from the diffuse astrophysical neutrino flux for this class of mergers.

In order to find the constraint on \f\ from the direct search, we calculate the expected neutrino flux that would be detectable on Earth, coming from GW150914, as a function of \f. We assume that the neutrinos are emitted with an $E^{-2}$-spectrum from a distance of 410~Mpc (corresponding to the central value of the LIGO estimated distance of 410$^{+160}_{-180}$~Mpc ). In order to convert the neutrino flux at Earth to the flux detected by IceCube, we use the effective area corresponding to the follow-up search~\cite{Aartsen:2014cva} in the declination band $-30^\circ < \delta < 0^\circ$. We include the irreducible background from atmospheric neutrinos~\cite{Sinegovskaya:2013wgm}. This background is integrated over a time interval of 1000~s, so that the signal is contained, and over a sky patch of 600~deg$^2$, which corresponds to the localization of GW150914 by LIGO.

\begin{figure}
	\centering
	\includegraphics[width=\figwidth\linewidth]{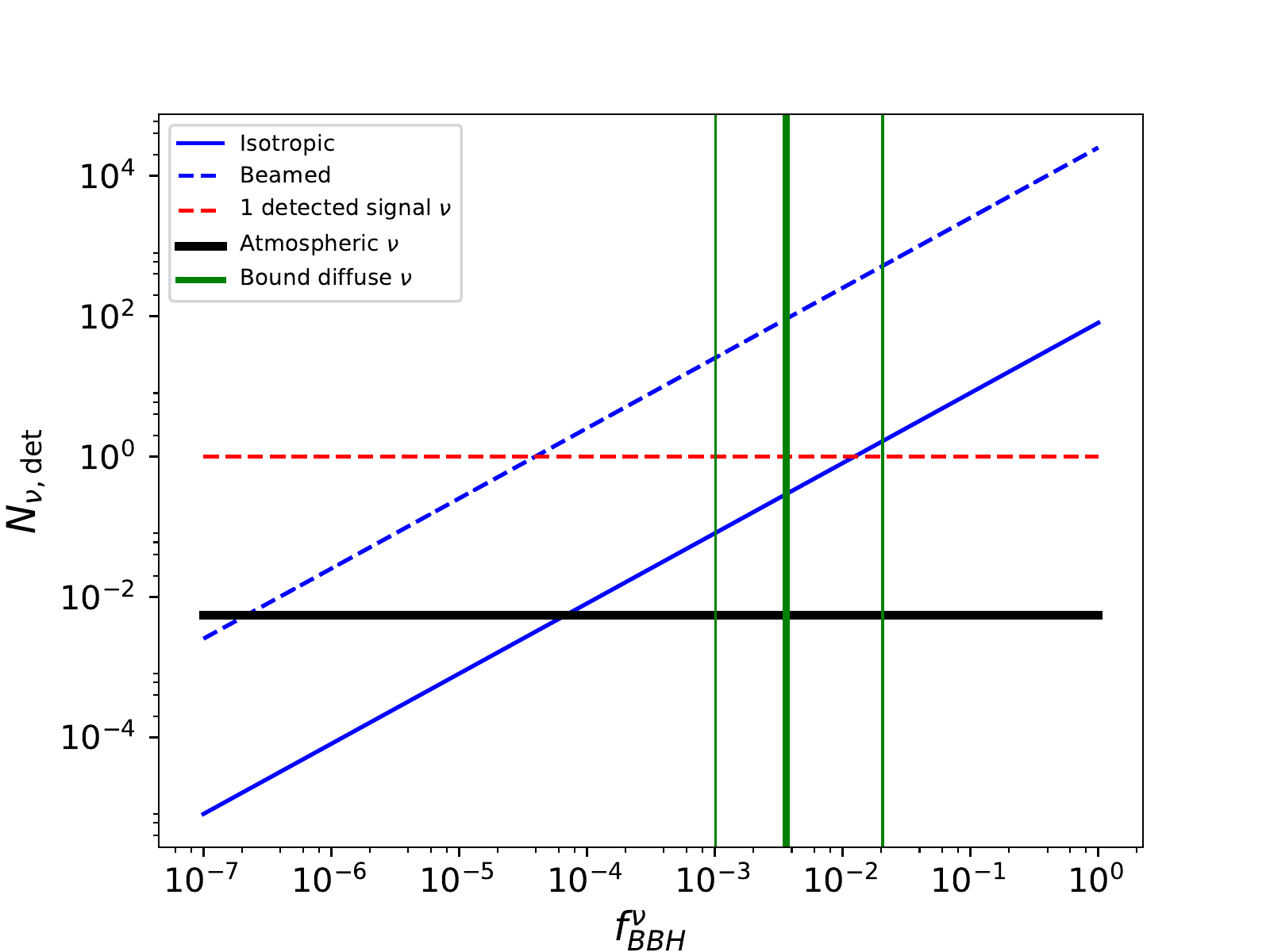}
	\caption{The expected number of neutrino events from GW150914, assuming an $E^{-2}$ neutrino spectrum, as a function of \f, for isotropic (full blue) and beamed (dashed blue) emission. The background, integrated over \SI{1000}{s} and a solid angle of 600~deg$^2$, is shown by the black line. The red dashed line indicates 1 detected neutrino. The limits on \f\ follows from the intersection of this line with the blue lines. The central value and uncertainty of the constraint from the astrophysical neutrino flux are shown in green. Figure from~\cite{deVries:2016ljw}.}
	\label{fig:integratedE2events}
\end{figure}

\figref{fig:integratedE2events} shows the expected number of detected neutrinos as a function of \f, for both isotropic emission (full blue line) and beamed emission within a solid angle of $0.2 \times 0.2$ (dashed blue line). From the non-detection of signal neutrinos, \f\ is constrained to be below the level where one signal neutrino would be detected (red dashed line). This corresponds to $\f= \num{1.24e-02}$ for isotropic emission and  $\f= \num{3.96e-05}$ for beamed emission. The irreducible background of atmospheric neutrinos is indicated by the solid black line. For a single BBH merger, this background is far below the level where one signal neutrino could be detected and is thus negligible. In order to calculate the bound from the diffuse astrophysical neutrino flux, we only considered the class of BBH mergers similar to GW150914, occurring with the rate in Eq.~\ref{eq:rate3m}. This bound is given by $\f=\numrange{1.01e-03}{2.06e-02}$, with a central value at $\f = \num{3.63e-03}$ (green lines).



\section{Prospects}

As more BBH mergers are detected, it will be possible to improve the limit on the neutrino emission by performing a stacked analysis. It will be assumed that all BBH mergers are similar to GW150914, injecting \egw of energy in gravitational waves and $\f \times \egw$ of energy into neutrinos from a distance of 410~Mpc. Only the case of isotropic emission will be shown, while the results for beamed emission can be immediately obtained by rescaling with the solid angle ${\frac{4\pi}{\Delta\Omega}=\frac{4\pi}{0.2\times0.2}}$. The analysis from the previous section is repeated, with an effective area averaged over the full sky. Since the localization of GW events is expected to improve in future observation runs~\cite{Gehrels:2015uga}, the bound on \f\ is calculated for background integrated over sky patches of 600~deg$^2$, 100~deg$^2$ and 20~deg$^2$. Because we now consider multiple events, the astrophysical bound is calculated with the full rate estimate in Eq.~\ref{eq:LIGOrate}.

\begin{figure}
	\centering
	\includegraphics[width=\figwidth\linewidth]{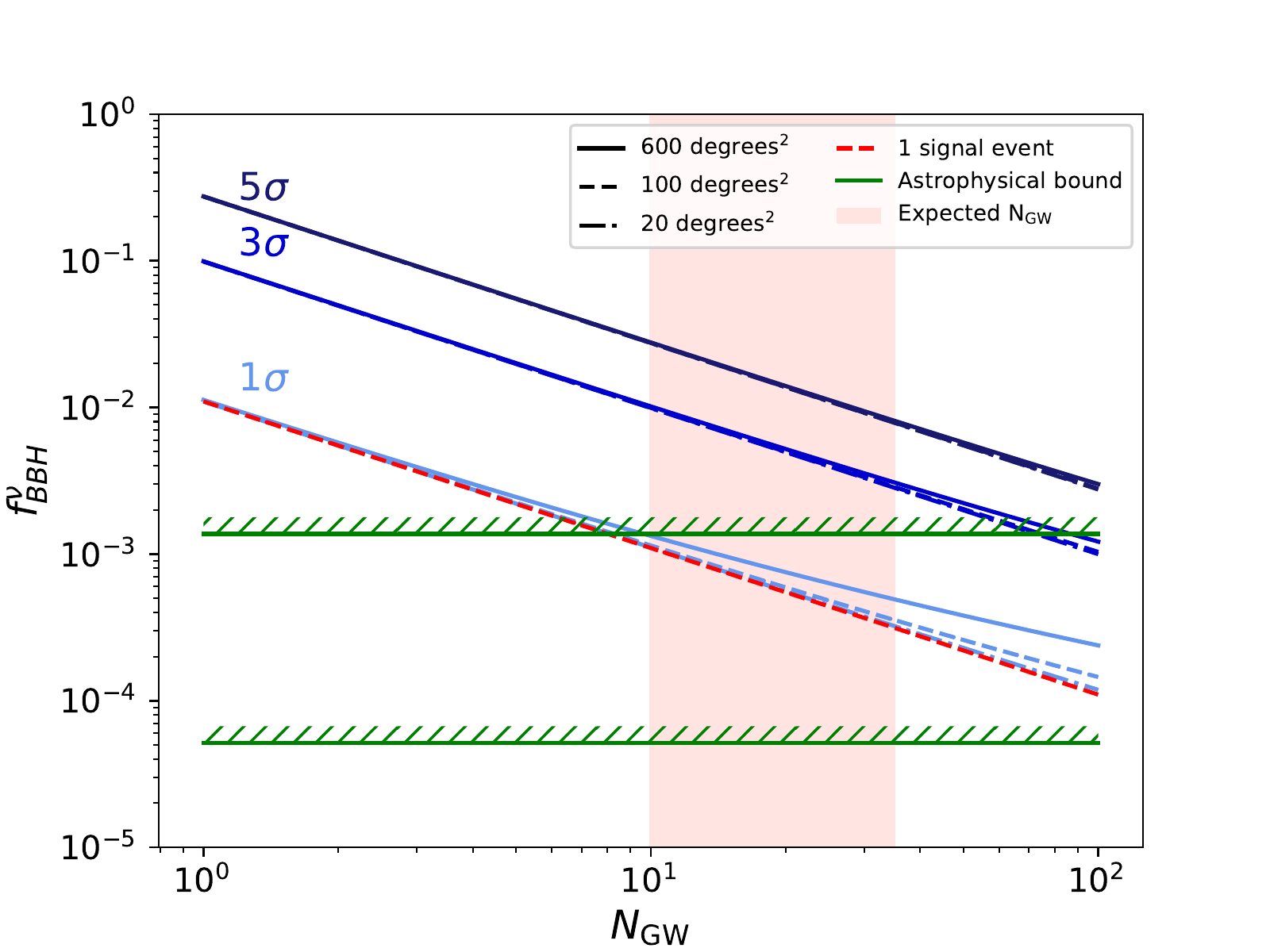}
	\caption{The expected limit on \f\ for different signal strengths (from top to bottom $S/\sqrt{S + B}=5,\,3,\,1$) as a function of number of \egw BBH mergers observed in gravitational waves. The background is integrated over a sky patch of 600~deg$^2$, 100~deg$^2$ and 20~deg$^2$ for the full, dashed and dashed-dotted lines respectively. The astrophysical bounds corresponding to the rates in Eq.~\protect\ref{eq:LIGOrate} are shown by the green hatched lines. The band indicates the expected number of BBH mergers in LIGO run O2. Figure from~\cite{deVries:2016ljw}.
}
	\label{fig:fracpotential}
\end{figure}

\figref{fig:fracpotential} shows the expected limit on \f\ as a function of the number of detected BBH mergers ($N_{\mathrm{GW}}$) by LIGO. The limit on \f\ from the stacked search with a signal strength $S/\sqrt{S + B}=5,\,3,\,1$ is given by the blue lines (from top to bottom). The level at which one signal neutrino would be detected in total is indicated by the red dashed line. The $S/\sqrt{S + B}=1$ limit follows this line up until 10 detected BBH mergers. Afterwards, the background starts to become important. The improvement in \f\ with localization by LIGO is indicated by the dashed and dashed-dotted blue lines. The expected number of BBH mergers in run O2 of LIGO is in the range \numrange{10}{35}~\cite{Abbott:2016nhf}. For the highest value of this range, it is possible to probe \f\ down to about $\f\approx \num{8e-3},\,\num{3e-3},\,\num{5e-4}$ for $S/\sqrt{S + B}=5,\,3,\,1$ respectively. The astrophysical bounds corresponding to the merger rates in Eq.~\ref{eq:LIGOrate} are at $\f = \num{1.37e-03}$ and $\f = \num{5.15e-05}$. The constraint on \f\ reaches the weakest astrophysical bound at $N_{\mathrm{GW}}\gtrsim 180,\,70,\,8$ for $S/\sqrt{S + B}=5,\,3,\,1$. A neutrino signal before 10 mergers would therefore violate the astrophysical bound, invalidating the assumptions in its calculation. This would therefore constrain the cosmic evolution of the BBH mergers or require the neutrino emission fraction \f\ to be different for certain BBH merger classes. If a neutrino signal is only detected for higher $N_{\mathrm{GW}}$, this will determine the level at which BBH mergers contribute to the diffuse astrophysical neutrino flux.

\section{Model-dependent interpretation}

It is useful to interpret this general limit on the neutrino emission fraction \f\ in more specific models of neutrino production. If it is assumed that no energy is extracted from the black holes or the gravitational waves themselves, but instead the neutrino emission originates from matter around the black hole binary, we can decompose \f\ as
\begin{equation}
\f = f_{\mathrm{matter}}\times f_{\mathrm{engine}} \times \epsilon_{\mathrm{p, acc}} \times \epsilon_{\nu}.
\label{eq:fparts}
\end{equation}
The first part, $f_{\mathrm{matter}}$, characterizes the amount of matter present in the black hole binary environment relative to the energy emitted in gravitational waves. The second part, $f_{\mathrm{engine}} \times \epsilon_{\mathrm{p, acc}} \times \epsilon_{\nu}$, characterises the acceleration of protons in order to create high energy neutrinos. The fraction of energy deposited from the surrounding matter in the acceleration engine is given by $f_{\mathrm{engine}}$. The amount of energy going from the engine to the protons is given by $\epsilon_{\mathrm{p, acc}}$, while the amount of energy going from protons into neutrinos is given by $\epsilon_{\nu}$.

If we assume that the acceleration of particles is similar to the case of a short GRB, we can use the GRB-fireball model~\cite{Waxman:1997ti} to calculate the acceleration part of Eq.~\ref{eq:fparts}. In this model, one typically finds that the accretion disk mass is converted into fireball energy with an efficiency  $f_{\mathrm{engine}}=1/10$. The protons can receive a fraction $\epsilon_{\mathrm{p, acc}}=1/10$ of the fireball energy and then produce neutrinos with $\epsilon_\nu=1/20$ of that energy. Therefore, we find that 
\begin{equation}
\f = f_{\mathrm{matter}}\times 5\cdot10^{-4}.
\end{equation}
This result can be used to put a bound on the amount of matter present in the binary black hole environment using the non-detection of counterpart neutrinos in GW150914. Since the bound in Section~\ref{section:direct} was at $\f= \num{3.96e-05}$ for beamed emission in a typical solid angle $\Delta \Omega=0.2\times0.2$, we estimate
\begin{equation}
f_{\mathrm{matter}}^{GW150914} \lesssim \num{7.9e-2} \times \frac{\Delta \Omega}{0.2\times0.2}.
\end{equation}
Using the expected limit at $S/\sqrt{S + B}=1$ on \f\ after 35 BBH merger events detected by LIGO from the previous section, the expected limit on the amount of matter is
\begin{equation}
f_{\mathrm{matter}}^{N_\mathrm{GW}=35} \lesssim \num{3.2e-3} \times \frac{\Delta \Omega}{0.2\times0.2}.
\end{equation}
If instead the bound on \f\ from the diffuse astrophysical neutrino flux from mergers similar to GW150914 is used (Section~\ref{section:direct}), which is independent of beaming, we find
\begin{equation}
f_{\mathrm{matter}}^{\mathrm{astrophys}} \lesssim \num{2.3e-2}.
\end{equation}

It is also possible to reverse the argument and find the neutrino emission fraction \f\ for realistic models of the black hole binary environment. In case the emission is coming from a non-active accretion disk that is reactivated upon the merger~\cite{Perna:2016jqh}, one expects about \numrange{.e-3}{.e-4} M$_\odot$ of matter. The expected neutrino emission fraction is then
\begin{equation}
\f\approx\num{.e-7},
\end{equation}
which is still below the expected bound in the near future (Fig.~\ref{fig:fracpotential}).
\section{Conclusion}

Using the observation of the binary black hole merger GW150914 and its corresponding follow-up search for high energy neutrinos by IceCube, it is possible to put a limit on the neutrino emission from binary black hole mergers. We parametrise this emission with the neutrino emission fraction $\f=\frac{E_\nu}{E_\mathrm{GW}}$. Using the non-detection of counterpart neutrinos in this event, it is possible to put a limit at $\f= \num{1.24e-02}$ for isotropic emission and  $\f= \num{3.96e-05}$ for beamed emission in a patch $\Delta\Omega=0.2\times0.2$. In addition, the observation of the diffuse astrophysical neutrino flux by IceCube limits the neutrino emission to $\f = \num{3.63e-03}$ (central value) from a class of mergers similar to GW150914.

The prospects for the expected limit on \f\ were shown as a function of the number of future BBH mergers observed by LIGO. It is found that after about 10 mergers, the limit on \f\ from direct searches will reach the limit set by the diffuse astrophysical neutrino flux at $S/\sqrt{S + B}=1$. The expected number of events in LIGO run O2 is between \numrange{10}{35}. Using the higher value of this estimate, we find that the expected limit of \f will reach $\f\approx \num{8e-3},\,\num{3e-3},\,\num{5e-4}$ for $S/\sqrt{S + B}=5,\,3,\,1$.

Finally, we show how to interpret this generic bound on the neutrino emission fraction \f\ in the context of specific models. Using the GRB-fireball model as the acceleration engine, we can constrain the amount of matter in the BBH merger environment to
${f_{\mathrm{matter}}^{GW150914} \lesssim \num{7.9e-2} \times \frac{\Delta \Omega}{0.2\times0.2}}$ from GW159014, $f_{\mathrm{matter}}^{N_\mathrm{GW}=35} \lesssim \num{3.2e-3} \times \frac{\Delta \Omega}{0.2\times0.2}$ from the expected limit on \f\ after 35 events and $f_{\mathrm{matter}}^{\mathrm{astrophys}} \lesssim \num{2.3e-2}$ from the astrophysical bound for mergers similar to GW150914. In order to give an estimate of \f\ for realistic situations, we applied the same method to a model of a dead accretion disk around one of the black holes which is reactivated upon the merger. We then find $\f\approx\num{.e-7}$, which is still below the expected reach in LIGO run O2.

\acknowledgments
KDV is supported by the Flemish Foundation for Scientific Research (FWO-12L3715N - K.D. de Vries). 
GDW and JMF are supported by Belgian Science Policy (IAP VII/37) and JMF is also supported in part by IISN. MV is aspirant FWO Vlaanderen. 

\bibliographystyle{unsrt}
\bibliography{biblio}

\begin{thebibliography}{10}

\bibitem{Abbott:2016blz}
B.~P. Abbott et~al.
\newblock {Observation of Gravitational Waves from a Binary Black Hole Merger}.
\newblock {\em Phys. Rev. Lett.}, 116(6):061102, 2016.

\bibitem{Perna:2016jqh}
Rosalba Perna, Davide Lazzati, and Bruno Giacomazzo.
\newblock {Short Gamma-Ray Bursts from the Merger of Two Black Holes}.
\newblock {\em Astrophys. J.}, 821(1):L18, 2016.

\bibitem{Abbott:2016gcq}
B.~P. Abbott et~al.
\newblock {Localization and broadband follow-up of the gravitational-wave
  transient GW150914}.
\newblock {\em Astrophys. J.}, 826(1):L13, 2016.

\bibitem{Adrian-Martinez:2016xgn}
S.~Adrian-Martinez et~al.
\newblock {High-energy Neutrino follow-up search of Gravitational Wave Event
  GW150914 with ANTARES and IceCube}.
\newblock {\em Phys. Rev.}, D93(12):122010, 2016.

\bibitem{Abe:2016jwn}
K.~Abe et~al.
\newblock {Search for Neutrinos in Super-Kamiokande associated with
  Gravitational Wave Events GW150914 and GW151226}.
\newblock {\em Astrophys. J.}, 830(1):L11, 2016.

\bibitem{Aab:2016ras}
Alexander Aab et~al.
\newblock {Ultrahigh-energy neutrino follow-up of Gravitational Wave events
  GW150914 and GW151226 with the Pierre Auger Observatory}.
\newblock {\em Submitted to: Phys. Rev. D}, 2016.

\bibitem{Connaughton:2016umz}
V.~Connaughton et~al.
\newblock {Fermi GBM Observations of LIGO Gravitational Wave event GW150914}.
\newblock {\em Astrophys. J.}, 826(1):L6, 2016.

\bibitem{Savchenko:2016kiv}
V.~Savchenko et~al.
\newblock {INTEGRAL upper limits on gamma-ray emission associated with the
  gravitational wave event GW150914}.
\newblock {\em Astrophys. J.}, 820(2):L36, 2016.

\bibitem{Bartos:2016dgn}
Imre Bartos, Bence Kocsis, Zoltán Haiman, and Szabolcs Márka.
\newblock {Rapid and Bright Stellar-mass Binary Black Hole Mergers in Active
  Galactic Nuclei}.
\newblock {\em Astrophys. J.}, 835(2):165, 2017.

\bibitem{Kotera:2016dmp}
Kumiko Kotera and Joseph Silk.
\newblock {Ultrahigh Energy Cosmic Rays and Black Hole Mergers}.
\newblock {\em Astrophys. J.}, 823(2):L29, 2016.

\bibitem{Aartsen:2016nxy}
M.~G. Aartsen et~al.
\newblock {The IceCube Neutrino Observatory: Instrumentation and Online
  Systems}.
\newblock {\em JINST}, 12(03):P03012, 2017.

\bibitem{deVries:2016ljw}
Krijn~D. de~Vries, Gwenha{\"e}l de~Wasseige, Jean-Marie Fr{\`e}re, and Matthias
  Vereecken.
\newblock {Constraints and prospects on GW and neutrino emissions using
  GW150914}.
\newblock 2016.

\bibitem{Aartsen:2013jdh}
M.~G. Aartsen et~al.
\newblock {Evidence for High-Energy Extraterrestrial Neutrinos at the IceCube
  Detector}.
\newblock {\em Science}, 342:1242856, 2013.

\bibitem{Kowalski:2014zda}
Marek Kowalski.
\newblock {Status of High-Energy Neutrino Astronomy}.
\newblock {\em J. Phys. Conf. Ser.}, 632(1):012039, 2015.

\bibitem{Ahlers:2014ioa}
Markus Ahlers and Francis Halzen.
\newblock {Pinpointing Extragalactic Neutrino Sources in Light of Recent
  IceCube Observations}.
\newblock {\em Phys. Rev.}, D90(4):043005, 2014.

\bibitem{Waxman:1998yy}
Eli Waxman and John~N. Bahcall.
\newblock {High-energy neutrinos from astrophysical sources: An Upper bound}.
\newblock {\em Phys. Rev.}, D59:023002, 1999.

\bibitem{Hopkins:2006bw}
Andrew~M. Hopkins and John~F. Beacom.
\newblock {On the normalisation of the cosmic star formation history}.
\newblock {\em Astrophys. J.}, 651:142--154, 2006.

\bibitem{Yuksel:2008cu}
Hasan Yuksel, Matthew~D. Kistler, John~F. Beacom, and Andrew~M. Hopkins.
\newblock {Revealing the High-Redshift Star Formation Rate with Gamma-Ray
  Bursts}.
\newblock {\em Astrophys. J.}, 683:L5--L8, 2008.

\bibitem{TheLIGOScientific:2016pea}
B.~P. Abbott et~al.
\newblock {Binary Black Hole Mergers in the first Advanced LIGO Observing Run}.
\newblock {\em Phys. Rev.}, X6(4):041015, 2016.

\bibitem{Aartsen:2015zva}
M.~G. Aartsen et~al.
\newblock {The IceCube Neutrino Observatory - Contributions to ICRC 2015 Part
  II: Atmospheric and Astrophysical Diffuse Neutrino Searches of All Flavors}.
\newblock In {\em {Proceedings, 34th International Cosmic Ray Conference (ICRC
  2015): The Hague, The Netherlands, July 30-August 6, 2015}}, 2015.

\bibitem{Aartsen:2014cva}
M.~G. Aartsen et~al.
\newblock {Searches for Extended and Point-like Neutrino Sources with Four
  Years of IceCube Data}.
\newblock {\em Astrophys. J.}, 796(2):109, 2014.

\bibitem{Sinegovskaya:2013wgm}
T.~S. Sinegovskaya, E.~V. Ogorodnikova, and S.~I. Sinegovsky.
\newblock {High-energy fluxes of atmospheric neutrinos}.
\newblock In {\em {Proceedings, 33rd International Cosmic Ray Conference
  (ICRC2013): Rio de Janeiro, Brazil, July 2-9, 2013}}, page 0040, 2013.

\bibitem{Gehrels:2015uga}
Neil Gehrels, John~K. Cannizzo, Jonah Kanner, Mansi~M. Kasliwal, Samaya
  Nissanke, and Leo~P. Singer.
\newblock {Galaxy Strategy for LIGO-Virgo Gravitational Wave Counterpart
  Searches}.
\newblock {\em Astrophys. J.}, 820(2):136, 2016.

\bibitem{Abbott:2016nhf}
B.~P. Abbott et~al.
\newblock {The Rate of Binary Black Hole Mergers Inferred from Advanced LIGO
  Observations Surrounding GW150914}.
\newblock {\em Astrophys. J.}, 833:1, 2016.

\bibitem{Waxman:1997ti}
Eli Waxman and John~N. Bahcall.
\newblock {High-energy neutrinos from cosmological gamma-ray burst fireballs}.
\newblock {\em Phys. Rev. Lett.}, 78:2292--2295, 1997.

\end{thebibliography}

\end{document}